\documentclass[aps,pra,twocolumn,showpacs,showkeys,reprint]{revtex4-1}
\usepackage{graphics}
\usepackage{graphicx}
\usepackage{amsmath}

\begin{document}

\title{Critical points of the anyon-Hubbard model}
\author{J. Arcila-Forero}
\author{R. Franco}
\author{J. Silva-Valencia}
\email{jsilvav@unal.edu.co}
\affiliation{Departamento de F\'{\i}sica, Universidad Nacional de Colombia, A. A. 5997 Bogot\'a, Colombia.}

\date{\today}

\begin{abstract}
Anyons are particles with fractional statistics that exhibit a nontrivial change in the wavefunction under an exchange of particles. Anyons can be 
considered to be a general category of particles that interpolate between fermions and bosons. We determined the position of the critical points of the 
one-dimensional anyon-Hubbard model, which was mapped to a modified Bose-Hubbard model where the tunneling depends on the local density and the 
interchange angle. We studied the latter model by using the density matrix renormalization group method  and observed that gapped (Mott insulator) and 
gapless (superfluid) phases characterized the phase diagram, regardless of the value of the statistical angle. The phase diagram for higher densities
was calculated and showed that the Mott lobes increase (decrease) as a function of the statistical angle (global density). The position of the critical 
point separating the gapped and gapless phases was found using quantum information tools, namely the block von Neumann entropy. We also studied 
the evolution of the critical point with the global density and the statistical angle and showed that the anyon-Hubbard model with a statistical 
angle $\theta =\pi/4$ is in the same universality class as the Bose-Hubbard model with two body interactions.
\end{abstract}


\maketitle

\section{Introduction}
Physicists have proposed  a third class of particles with nontrivial exchange statistics, anyons, particles carrying fractional statistics that 
interpolate between bosons and fermions~\cite{Leinaas-NCB77,Wilckzek-PRL82,Haldane-PRL91}.  For two anyons under particle exchange, the wave 
function acquires a fractional phase $e^{i\theta} $ , giving rise to fractional statistics with $0<\theta <\pi$. Greater interest in the 
study of anyons emerged when the fractional quantum Hall effect, observed experimentally, had a natural explanation in terms of 
anyons~\cite{Tsui-PRL82,Laughlin-PRL83}. Another discovery that reinforced this interest was evidence of superconductor anyon 
gas~\cite{Fetter-PRB89,Chen-IJMP89}. Anyons are very important  in numerous studies related to the fractional quantum Hall 
effect~\cite{Halperin-PRL84,Camino-PRB05}, condensed matter physics, and topological quantum computation~\cite{Kitaev-AP03,Nayak-RMP08,Alicea-NP11}. 
The study of anyons was restricted for many years to two-dimensional systems. However, with Haldane's definition of fractional statistics, it 
was generalized to arbitrary dimensions~\cite{Haldane-PRL91}.\par
One-dimensional (1D) anyons have been studied from different theoretical approaches. Kundu obtained the exact solution of the one-dimensional 
anyon gas using the generalized coordinate Bethe ansatz method and found the generalized commutation relations for anyons~\cite{Kundu-PRL99}. 
Furthermore, Batchelor  {\it et al.} showed that the low energies, the dispersion relations, and the generalized exclusion statistics depend on both 
the anyonic statistical angle and the dynamical interaction parameters in a 1D anyon gas~\cite{Batchelor-PRL06}. Alternatively, in tight waveguides, 
the Fermi-Bose mapping method  for one-dimensional Bose and Fermi gases was generalized to an anyon-fermion mapping and applied in order to obtain 
exact solutions of several models of ultracold gases with anyonic exchange symmetry~\cite{Girardeau-PRL06}. In 2007, Calabrese and Mintchev studied 
the correlation functions of the 1D anyonic gapless systems in the low-momentum regime~\cite{Calabrese-PRB07}. Interesting features appear, including 
universal oscillating terms with frequency proportional to the statistical parameter and beating effects close to the fermion points. Later, 
Vitoriano and Coutinho-Filho~\cite{Vitoriano-PRL09} studied the ground state and low-temperature properties of an integrable Hubbard model with 
bond-charge interaction, finding that the model displays fractional statistical properties. Remarkably, one-dimensional anyons can be realized as 
low-energy excitations of the Hubbard model of fermions with correlated hopping processes. On the other hand, Hao {\it et al.} investigated 
the ground state~\cite{Hao-PRA09} and dynamical~\cite{Hao-PRA12} properties of anyons confined in one-dimensional optical lattices with a 
weak harmonic trap using an exact numerical method based on a generalized Jordan-Wigner transformation.
Also, two-component mixtures of anyons under an external trap were considered by Zinner~\cite{Zinner-PRA15} and the correlation functions of 
one-dimensional hard-core anyons were calculated by Patu~\cite{Patu-JSM15}.\par
Note that various experimental proposals for the creation, detection, and manipulation of anyons have been made. Rotating Bose-Einstein 
condensates have been used to create anyons~\cite{Paredes-PRL01}, and the results can be understood in terms of the fractional quantum Hall effect 
for bosons~\cite{Xie-PRL91}. This system offers the formation of particles exhibiting fractional statistics with a well-controlled setup that can 
allow experimentalists to test their fractional statistics. Later on, Duan and colleagues described a general technique for controlling many-body 
spin Hamiltonians using ultracold atoms, and they showed how to implement an exactly-solvable spin Hamiltonian that supports Abelian and 
non-Abelian anyonic excitations with exotic fractional statistics~\cite{Duan-PRL03}. On the other hand, it is possible to use an atomic spin lattice 
in optical cavities for the direct measurement of anyonic statistics~\cite{Jiang-NP08} or trapped atoms in an optical lattice in order to create 
anyons in topological lattice models. These types of schemes allow the creation of topologically ordered states and detect their 
statistics~\cite{Aguado-PRL08}. Alternatively, a suggestion has been made for creating anyons on a 1D lattice based on light propagation in an 
engineered array of optical waveguides. This photonic setup enables us to see the impact of the statistical exchange phase $\theta$ on the 
correlated tunneling dynamics~\cite{Longhi-OL12}. Furthermore, the possibility of realizing the bosonic fractional quantum Hall effect in ultracold 
atomic systems has been shown, suggesting a new route to producing and manipulating anyons~\cite{Zhang-PRB15}.\par
We would especially like point out that several proposals for using ultracold bosons to produce anyons in an optical lattice have been made.  In 
particular, Keilmann {\it et al.} propose a realistic setup for demonstrating an interacting gas of anyons using Raman-assisted hopping in a 1D 
optical lattice~\cite{Keilmann-NC11}. 
They introduced the anyon-Hubbard model, which is equivalent to a modified  Bose-Hubbard model in which the bosonic hopping depends on the 
local density. This is an exact mapping between anyons and bosons in one dimension. They found, among other things, the phase diagram at 
zero-temperature with density $\rho=1$ using the density matrix renormalization group, and concluded that the anyons in 1D display insulator and 
superfluid phases. The first is characterized by having a nonzero energy gap in which the atoms are localized on the lattice, and the second 
by a gapless phase in which the atoms are delocalized and dispersed across the lattice. In addition, they presented the mean-field solution for 
the Mott-superfluid transition for different angles and a comparison with the bosonic case, where it is possible to see the expansion of the Mott 
lobes with the statistical angle. Later on, the ground-state properties of anyons in a one-dimensional 
lattice were analyzed by Tang {\it et al.}~\cite{Tang-NJP15} using the Hamiltonian proposed by Keilmann {\it et al.}~\cite{Keilmann-NC11}, and they 
obtained that anyons have an asymmetric quasi-momentum distribution, where the peak position depends on both the fractional phase and the particle 
number density.  In the same way, the momentum distributions and the effects of the statistical angle on the correlations were analyzed using the 
density matrix renormalization group and mean field methods by Zhang {\it et al.}, finding that the statistical angle could modulate the beat length 
of the correlations~\cite{Zhang-arxiv15}.\par 
In 2015, Greschner and Santos proposed an experimental scheme to improve the proposal for the realization of Keilmann's anyon-Hubbard model. This 
scheme allows as well for an exact realization of the two-body hard constraint ( {\it i.e.} $ (b^{ \dagger}_{j})^3 =0$) and  controllable effective 
interactions without the need for Feshbach resonances. They show that the interplay of anyonic statistics, two-body hard constraint, and controllable 
interactions results in a far richer physics for the model, and the phase diagram includes a pair-superfluid, a dimer, and an exotic partially-paired 
phase~\cite{Greschner-PRL15}.\par
This year, a simple scheme for realizing the physics of 1D anyons with ultracold bosonic atoms in an optical lattice has been elaborated. It 
relies on lattice-shaking-induced resonant tunneling against potential off-sets created by a combination of a lattice tilt and strong on-site 
interactions. No lasers in addition to those used for the creation of the optical lattice are required~\cite{Strater-ArX2016}. Also, they 
calculated the density and the second Renyi entropy of a chain with twenty sites, using exact diagonalization.\par
Taking into account the above facts, it is clear that the anyon-Hubbard model is a very exciting problem, and we want to contribute to it. 
From the previous results, we know that the insulator regions grow as the statistical angle $\theta$ increases for density $\rho=1$. In the present 
case, we are interested in the study of the influence of the increase in the density of the system on the position of the point at which the two 
phases separate. Furthermore, the mean-field solution for $\theta=\pi/4$~\cite{Keilmann-NC11} indicates a possible odd-even asymmetry in which the 
insulator region with an even density increases in comparison with the lobes with odd densities. Hence we will study the system properties with 
$\theta=\pi/4$  to verify or challenge the mean-field results.\par 
In this paper, we study the anyon-Hubbard model using the density matrix renormalization group (DMRG) method. Using the energy for adding and removing 
particles in the system, we construct the phase diagram for $\theta=\pi/4$ in the plane ($t/U$, $\mu/U$) for three densities ($\rho=1,2$, and $3$) 
and we conclude that as we increase the density, the position of the critical point changes to lower values of kinetic energy $t/U$. These 
results contrast with previous mean-field calculations for $\theta=\pi/4$~\cite{Keilmann-NC11}. On the other hand, it is necessary to find a tool that 
gives us a better estimate of the critical point than simply the gap closing. We consider that during the last decades, it has been shown that quantum 
information tools are useful for studying the border between the quantum phases~\cite{Amico-RMP08}. In this system, the phase transition was 
studied using the block von Neumman entropy, and we were able to observe 
a critical and a noncritical behavior in the system related to the superfluid to Mott insulator transition. In particular, we use 
the estimator proposed by L{\"a}uchli and Kollath~\cite{Lauchli-JSM08} to determine the critical points of the anyon-Hubbard model. We obtain 
the evolution of the critical points with the density and find functional relationships between the different 
parameters that the Hamiltonian depends on. We show that a simple analytical function is a good approximation of the results. It is important to 
note that studies related to the most precise estimation of critical points have not been previously reported, beyond the gap closing. We use the 
critical points and show that the Kosterlitz-Thouless formula is suitable for describing the closing of the gap, and we can infer that the 
anyon-Hubbard model with $\theta=\pi/4$ is in the same universality class as the Bose-Hubbard model.\par

The outline of this paper is as follows: In Sec. \ref{ii}  we introduce the anyon-Hubbard model. In sec. \ref{iii}  we show the phase diagram 
obtained and we study the critical points of the system through the von Neumann block entropy. Finally, in Sec. \ref{iv} we 
summarize our results.

\subsection{\label{ii} Model}

The anyon-Hubbard model takes into account the hopping of the anyons along the lattice and the local two-body interaction between 
them~\cite{Keilmann-NC11}, and its Hamiltonian is given by  
\begin{equation}
\label{aHmH}
H=-t\sum_j^{L-1}\left(     a_j^{\dag} a_{j+1}  + h.c   \right) + \frac{U}{2} \sum_j^{L} n_j(n_j - 1),
\end{equation}

\noindent where $t>0$ is the tunneling amplitude connecting two neighboring sites, $U$ is the on-site interaction, $L$ is the length of the 
lattice, $n_j$ is the number operator and the operators $a_j^{\dagger}$, and  $a_j$ creates or annihilates an anyon on site $j$, respectively. 
The above creation and destruction operators satisfy the following commutation relation:
\begin{eqnarray}
a_j a_k^{\dagger} - e^{-i\theta sgn(j-k)}a_k^{\dagger}a_j=\delta_{jk},\\\nonumber 
a_j a_k= e^{i\theta sgn(j-k)}a_ka_j, 
\label{crelan}
\end{eqnarray}
\noindent where $\theta$ denotes the statistical phase, and the sign function is $sgn(j-k)=\pm 1$ for $j>k$ and $j<k$, and $=0$ for $j=k$. Note that 
the bosonic commutation relations are reproduced at the same site ($j=k$). Thus two particles at the same site behave as ordinary bosons. In 
consequence, when $\theta=\pi$,  we have pseudofermions: in spite of being bosons on-site, they are fermions off-site.\par 
To study the ground state of the Hamiltonian (\ref{aHmH}), Keilmann {\it et al.} \cite{Keilmann-NC11} proposed an exact mapping between anyons and 
bosons in 1D by means of the fractional version of a Jordan-Wigner transformation given by: 
$a_j=b_j exp\left( i\theta \sum_{i=1}^{j-1} n_i \right) \label{maping}$, 
where the operator $b_j$ describes spinless bosons, which satisfy $[b_j,b_i^{\dagger}]=\delta_{ji}$ and $[b_j,b_i]=0$. The number operator is defined 
by $n_i=a_i^{\dagger}a_i=b_i^{\dagger}b_i$.\par 
After using the anyon-boson mapping, the anyon-Hubbard Hamitonian is given  in terms of bosonic operators thus:
\begin{eqnarray}
\label{hahbo}
H&=&-t\sum_j^{L-1}\left(     b_j^{\dag} b_{j+1} e^{i\theta n_j} + h.c.   \right) \\\nonumber
&&+ \hspace{0.2 cm}  \frac{U}{2} \sum_j^{L} n_j(n_j - 1).\\\nonumber
\end{eqnarray}
Note that the above Hamiltonian describes bosons with an occupation-dependent amplitude $te^{i\theta n_j}$ for hopping processes from right to 
left ($j+1\rightarrow j$). If the target site $j$ is unoccupied, the hopping amplitude is simply $t$. If it is occupied by one boson, the amplitude 
reads $te^{i\theta}$, for two bosons $te^{i2\theta}$, and so on. We fix the energy scales by considering $U=1$. \par
It is important to observe that, for $\theta=0$ the anyon commutation relations revert to the well-known bosonic relations, and the anyon-Hubbard 
model corresponds to the well-known Bose-Hubbard model in this limit. Many analytical and numerical approaches have been used to study the ground 
state of the Bose-Hubbard model, and we know that for large $t$, the bosons would be completely delocalized in the lattice and the system would be in 
a superfluid state. When $U$ dominates, an integer number of bosons would be localized at each site, and the ground state is a Mott insulator one. 
The border between the superfluid and the Mott insulator regions can be estimated with the energy for adding and removing particles:

\begin{eqnarray}
\mu^{p}(L)=E_0 (L, N+1)-E_0 (L, N),\label{part}\\
\mu^{h}(L)=E_0 (L, N)-E_0 (L, N-1),\label{hole}\\\nonumber
\end{eqnarray} 

\noindent where $E_0 (L, N)$ denotes the ground-state energy for $L$ sites and $N$ particles. If the above parameters ($L$ and $N$) are finite, we 
observe that the single-particle excitations exhibit a finite gap $\Delta\mu(L)=\mu^{p}(L)-\mu^{h}(L)=E_0 (L, N+1)+E_0 (L, N-1) - 2E_0 (L,N) $. A 
Mott insulator state is achieved if the density of the system $\rho=N/L$ is an integer and is at the thermodynamic limit 
$\Delta\mu=lim_{L,N\rightarrow \infty} \Delta\mu(L)>0$. By contrast, the superfluid phase is gapless.\par 

\begin{figure}[t]
\setlength{\fboxrule}{0.5 pt}
\includegraphics [width=0.52\textwidth]{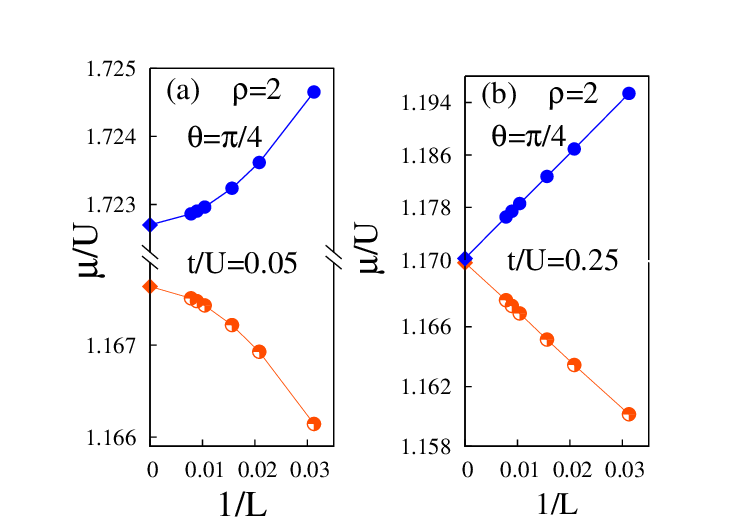} 
\caption{(Color online) System size dependence of the chemical potential of anyons in 1D with statistical angle $\theta=\pi/4$ and $\rho=2$. The upper set of data  
in each panel corresponds to the particle excitation energy and the lower one to the hole excitation energy. In the left panel ($t/U=0.05$),
we show a state with a finite difference at the thermodynamic limit, while this difference vanishes in the right panel($t/U=0.25$).}
\label{fig1}
\end{figure}
\subsection{\label{iii}  Results}
To calculate the ground state of a lattice with $L$ sites and $N$ particles, we truncated the local Hilbert space by considering only $\rho+5$ states 
when the density of the particles is $\rho$~\cite{Carrasquilla-PRA13} and used the finite-size density matrix renormalization group algorithm (DMRG) 
with open boundary conditions. Also, we used the dynamical block state selection (DBSS) protocol based on a fixed truncation error of the subsystem's 
reduced density matrix instead of using a fixed number of preserved states in the DMRG sweeps~\cite{Legeza-PRB03}. Using this protocol, we obtained a 
discarded weight of around $10^{-9}$ or less, and the maximum number of states retained was $m=1080$.\par
In the context of the quantum Hall regime, an experimental setup with a superconducting film adjacent to a two-dimensional electron gas can be 
understood in terms of anyons with a statistical angle $\theta=\pi/4$, and this system could prove useful in schemes for fault-tolerant topological 
quantum computation~\cite{Weeks-NP07}. A mean-field calculation of the phase diagram of one-dimensional anyons for densities $\rho=1,2$, and $3$  was 
presented byKeilmann et al. for $\theta=\pi/4$~\cite{Keilmann-NC11}. The above facts motivated us to consider this special angle in the first part of 
our study.\par
\begin{figure}[t]
  \centering
\setlength{\fboxrule}{0.5 pt}
 \includegraphics [width=0.5\textwidth]{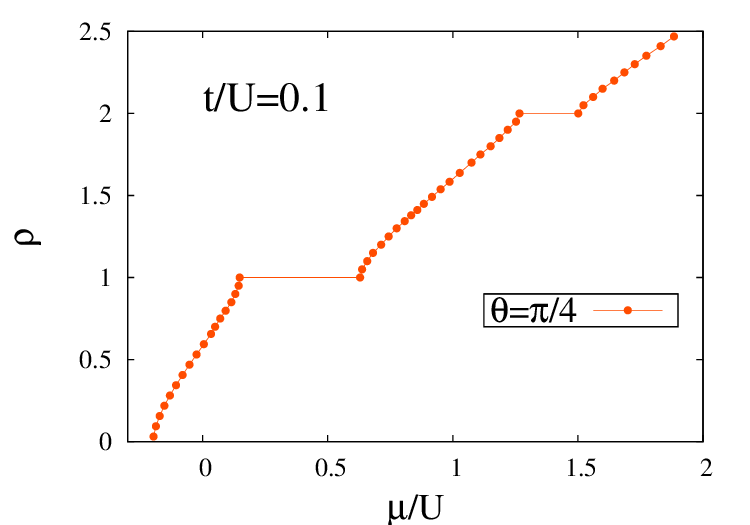} 
 \caption{  (Color online) Density $\rho$ versus chemical potential $\mu$ at the thermodynamic limit for $t/U=0.1$ and a statistical 
 angle $\theta=\pi/4$. }
\label{fig2}
\end{figure}
The evolution of the energies for adding and removing particles given by Eq. \ref{part} and Eq. \ref{hole} versus the inverse of the lattice length for anyons 
with $\theta=\pi/4$ and density $\rho=2$ appear in Fig. \ref{fig1}. In each panel, the upper (lower) curve corresponds to the energy for adding 
(removing) particles. Regardless of the value of the hopping parameter, we obtained that the energy for adding (removing) particles always decreases 
(increases) as a function of $1/L$; however, this evolution is quadratic for $t/U=0.05$, and at the thermodynamic limit we obtain 
$\Delta\mu/U=lim_{L,N\rightarrow \infty} [\mu^p(L)-\mu^h(L) ]=0.55$, which suggests that the ground state corresponds to a Mott insulator one. On the 
other hand, for $t/U=0.25$, the evolution is linear, the energy for adding and removing particles meets at the thermodynamic limit, and the ground 
state is superfluid. This figure tells us that the anyon liquid passes from a Mott insulator state to a superfluid one when the kinetic energy 
increases; hence its behavior is similar to the Bose-Hubbard model ($\theta=0$), and the main difference will be the position of the critical 
point.\par
In Fig. \ref{fig2}, we show the density $\rho$ as a function of the chemical potential, which was found at the thermodynamic limit value. We observe 
that the chemical potential increases as the density grows; however this behavior changes when the density reaches integer values. For $\rho=1$ and 
$\rho=2$, we obtain two plateaus in the curve, which indicates that the ground state has a finite gap for integer densities, whereas the width of these 
plateaus give us the value of the gap. Comparing this with Fig. \ref{fig1}, we obtain that for a ground state with two bosons per site, the gap will 
decrease monotonously as the hopping parameter increases. An important fact in Fig. \ref{fig2} is that the slope is always greater than zero, i. e. 
the compressibility $\kappa=\partial\rho/\partial\mu >0$, an argument that is related to the absence of first-order transition. Note that Batrouni and his
collaborators have shown the existence of first-order phase transitions  ($\kappa <0$) in two-dimensional system of 
spinless~\cite{Batrouni-PRL00} and spinor~\cite{Forges-PRB13} bosons, but for even lobes in one-dimensional systems of spin-1 bosons,
the compressibility is positive and the phase transition is of first order, this being caused by the spin degree of freedom~\cite{Batrouni-PRL09}. 
Taking into account the above discussion and our numerical results, we believe that the phase transitions for $\theta=\pi/4$ are of the second order kind; 
however the possibility of find first-order transitions in the anyon-Hubbard model for larger values of $\theta$  and/or a fixed number of particles 
is an interesting open problem.\par 
\begin{figure}[t]
  \centering
\setlength{\fboxrule}{0.5 pt}
 \includegraphics [width=0.5\textwidth]{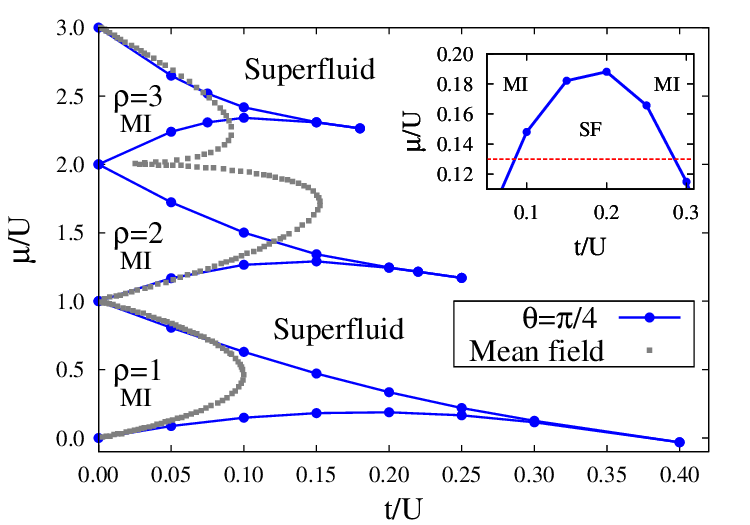} 
 \caption{  (Color online) Phase diagram of the anyon-Hubbard model with statistical angle $\theta=\pi/4$ for the densities $\rho=1,2$, and $3$ using 
 DMRG (blue line-circle) and comparison with the mean-field solution (gray squares) for the same densities (mean-field data were taken 
 from \cite{Keilmann-NC11}). Inset: display sequence from Mott insulator to supefluid and back to Mott insulator for $\theta=\pi/4$ and $\rho=1$ at 
 fixed $\mu=0.13$.}
\label{fig3}
\end{figure}
The mean-field phase diagram of Hamiltonian (\ref{aHmH}) found by Keilmann {\it et al.} \cite{Keilmann-NC11} is reproduced in Fig. \ref{fig3}
(gray squares). For $\theta=\pi/4$, we note that the Mott insulator lobes are surrounded by the superfluid phase, and their shapes are rounded. 
The mean-field solution shows that the critical points for all densities are lower than the ones found for the Bose-Hubbard model, and that the 
second lobe is larger than the other lobes. This result suggests a possible odd-even asymmetry present in the system in which lobes with an even 
density ($\rho=2$) increase in comparison with those with odd densities ($\rho=1$ and $3$). In addition, we can see that the critical point for the 
densities $\rho=1$ and $\rho=3$ do not differ considerably.\par 
Despite the interesting results of the mean-field solution of Hamiltonian (\ref{aHmH}), a determination of the phase diagram for higher densities 
($\rho>1$) beyond mean-field has not been done. For this reason, we calculated the chemical potential at the thermodynamic limit and found the phase 
diagram for $\theta=\pi/4$ and the three densities $\rho=1,2,$, and $3$ (blue line-circle) at the plane $(\mu/U,t/U)$. For small values of $t/U$, we 
see that the borders of the first Mott lobe obtained by mean-field and DMRG are closer; however for larger values the mean-field solution lobe closes, 
while the DMRG solution stretches slowly, and the gap closes at around $t/U=0.40$, which is four times larger than the mean-field result [see 
Fig. \ref{fig3}]. Note that our results are in agreement with the DMRG results of Keilmann {\it et al.} for $\rho=1$. For the second lobe, we observe 
that the mean-field and DMRG solutions only coincide at the lower edge for small values of $t/U$. The area of the mean-field solution is greater and 
the DMRG gap closes at around $t/U=0.25$, which indicates that there is no odd-even asymmetry between the lobes, this being an artifice of the 
mean-field solution. For the lobe with density $\rho=3$, we obtain that the upper borders of both solutions are closer for small values of $t/U$ and 
the critical point for this density is located around $t/U=0.18$, which is lower than the location of the critical points for the other densities.\par 
In Fig. \ref{fig3}, we observe that the  area of the Mott insulator lobes decreases as the global density of the system increases, which implies that 
the location of the critical points moves to lower values with the density. These facts are consistent with those obtained for the Bose-Hubbard 
model~\cite{Ejima-EL11}. However, the effect of anyonic statistics reflected in a correlated density dependent hopping is larger values of the 
critical points for all the densities considered, i. e., for a nonzero statistical angle, we need more kinetic energy to delocalize the particles and 
generate a superfluid state. Note that for all lobes, the gap closes as the hopping parameter increases and the shape of lobes becomes elongated, with 
a large tip, which indicates that the gap closes very  slowly, a fact is relevant to determination of the type of phase transition.\par
In the inset of Fig. \ref{fig3}, we show a zoom of the lower edge of the first lobe and a chemical potential constant line ($\mu=0.13$). Based on 
this inset, it is clear that the hole excitation energy has a maximum value, a fact repeated for the others lobes, but the position of the maximum 
moves to lower values of the hopping. Moving on along the red line, we observe that for small values of $t/U$ there is one boson per site, and the 
ground state is a Mott insulator one, but for bigger values a quantum phase transition takes place and the system passes to a superfluid phase with 
a global density lower than one. When  $t/U\approx0.28$, the system re-enters into a Mott insulator phase with density $\rho=1$, and a new transition 
to a superfluid phase with global density greater than one is expected for larger values of the hopping. Note that this reentrance phase transition 
also happens for the other lobes calculated here. This fact was first discussed by Kuhner et al. for the Bose-Hubbard model ($\theta=0$) with 
density $\rho=1$~\cite{Kuhner-PRB00}, and recently a detailed study was conducted by Pino et al. ~\cite{Pino-PRA12}.\par
From the above discussion and the previous papers about the anyon-Hubbard model, we know that the ground state exhibits a Mott insulator phase and a 
superfluid phase whose boundaries were found here for three different densities considering an angle of $\theta=\pi/4$. However, to use the closing 
gap criterion to determine the critical point when the density is fixed is not appropriate, as has been widely discussed for the Bose-Hubbard model. 
The precise determination of the critical points of this last-named model has received significant attention in the last decade, and many approaches 
have been considered, for instance using the interaction parameter of the Luttinger liquid~\cite{Kuhner-PRB00,Ejima-EL11}, or the tools of quantum 
information theory~\cite{Lauchli-JSM08,Rachel-PRL12,Ejima-PRA12}. Clearly, for the scientific community a precise determination of the critical 
points that separate the two quantum phases for the anyon-Hubbard model is a very interesting problem.\par
Today, entanglement is an important tool for studying the ground state of strongly correlated systems as well as the quantum phase transitions that 
will occur in the system. Measures of the entanglement, such as fidelity, von Neumann entropy, purity, and negativity, among others have been useful 
in determining the critical points of diverse models~\cite{Amico-RMP08}. In the present paper, we will use the von Neumann block entropy for studying 
the ground state of the Hamiltonian (\ref{hahbo}).\par
We consider a system with $L$ sites divided into two parts. Part $A$ has $l$ sites ($l=1,...,L$), and the rest form part $B$, with $L-l$ sites. 
The von Neumann block entropy of block $A$ is defined by $ S_A=-Tr\varrho_A  ln  \varrho_A$, where $\varrho_A=Tr_B \varrho$ is the reduced density 
matrix of block $A$ and $\varrho= |\Psi\rangle \langle \Psi|$ the pure-state density matrix of the whole system. For a system with open boundary 
conditions, the behavior of the von Neumann block entropy as a function of $l$ depends on the nature of ground state 
and provides information about the type of phase, because it saturates(diverges) if the system is gapped (gapless)~\cite{Calabrese-JSM04}, thus:

\begin{equation}
   S_L(l)= \left\{ \begin{array}{lcc}
             \frac{c}{6}ln [\frac{2 L}{\pi} sin (\frac{\pi l}{L})] + \Theta, & \text{critical}, \\
             \\ \frac{c}{6}ln[\xi_L]+\Theta^{\prime}, &\text{non critical},  
             \end{array}   
\right.
\label{lauchhh}
\end{equation}
\noindent where $c$ is the central charge and $\xi_L$ is the correlation length. The constants $\Theta$ and $\Theta^\prime$ are nonuniversal and model 
dependent.
\begin{figure}[t]
  \centering
\setlength{\fboxrule}{0.5 pt}
 \includegraphics [width=0.5\textwidth]{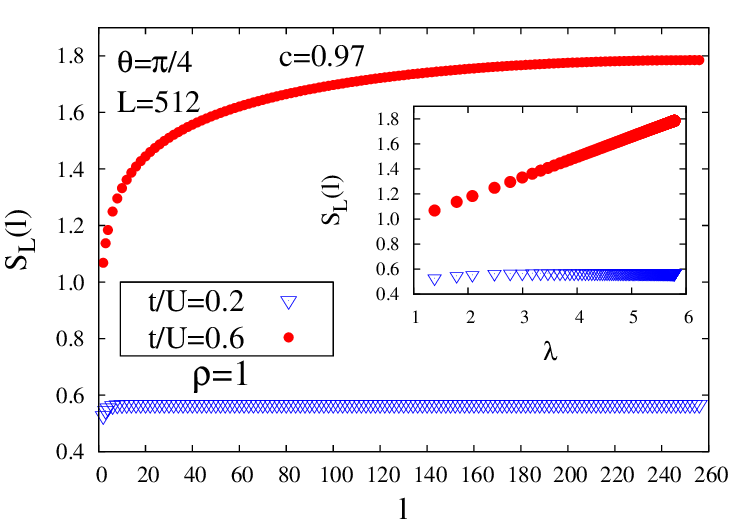} 
 \caption{ (Color online) The von Neumann block entropy $S_L(l)$ as a function of $l$ for a system with size $L=512$, $\rho=1$, and $\theta=\pi/4$. 
 Here we consider two different values of the hopping parameter, $t/U=0.2$ and $t/U=0.6$. In the inset, the von Neumann block entropy $S_L(l)$ as function 
 of the logarithmic conformal distance $\lambda$ is shown, revealing a linear behavior for the critical state. Otherwise, the non-critical state does not exhibit 
 linear behavior, because of the short correlation length. We found that the central charge is $c=0.97$.}
\label{fig4}
\end{figure}
The von Neumann block entropy $S_L(l)$ as a function of the block size $l$ is shown in Fig. \ref{fig4} for a lattice with global density $\rho=1$, 
statistical angle $\theta=\pi/4$, and two different values of the hopping: $t/U=0.2$ and $t/U=0.6$. At the limit $t\rightarrow 0$, the ground state 
can be seen as a product of local states, i. e., it is separable, and we expected that the entanglement would be zero. For a nonzero value of the 
hopping $t/U=0.2$, we observe that the block entropy is different from zero; it increases rapidly, and saturates at a certain value, in accordance 
with the expression Eq. (\ref{lauchhh}), which indicates that the ground state has a finite correlation length,  as is characteristic of the Mott 
insulator phase. A different behavior of $S_L(l)$ as a function of $l$ is observed for $t/U=0.6$; now the von Neumann block entropy always grows with 
the block size and diverges, which characterizes a critical state. In the inset of Fig. \ref{fig4}, we show the relationship between block entropy and 
the logarithmic conformal distance ($\lambda=ln \left[\frac{2L}{\pi}sin \left(\frac{\pi l}{L}\right)\right]$). We obtain a nonlinear dependence for 
small values of the hopping ($t/U=0.2$) due to the short correlation length, and linear behavior is observed for the critical state ($t/U=0.6$). 
Observing the expression (\ref{lauchhh}), we note that the slope of the block entropy versus the logarithmic conformal distance is related to the 
central charge of conformal theory; hence from the inset of Fig. \ref{fig4} we obtain the central charge $c=0.97$. This value is very close to 1, 
which corresponds to the central charge for the Bose-Hubbard model ($\theta=0$). Specifically, in the supefluid phase the low-energy physics of the 
one-dimensional Bose-Hubbard model can be described as a Luttinger liquid, which is a conformal field theory with central charge 
$c=1$~\cite{Lauchli-JSM08}.\par 

When the hopping increases from zero, the von Neumann block entropy allows us to identify two different ground states, one critical and the other not; 
however, identifying for which value of $t/U$ the transition takes place is not an easy task. Nevertheless, we can calculate the block entropy for different 
values of $t/U$ and try to estimate the critical value for which the system passes from a saturation behavior to a critical one, which could be a criterion for 
determining the critical point. In reality, this is very poor and needs a very large number of calculations. This problem was addressed by 
L{\"a}uchli and Kollath,  who proposed an estimator in terms of the von Neumann block entropy defined by the following expression: 
$\Delta S(L)= S_L(L/2)-S_{L/2}(L/4)$. This measures the increase of the entropy at the mid-system interface upon doubling the system size 
\cite{Lauchli-JSM08}.  According to (\ref{lauchhh}), we obtain: 

\begin{equation}
   \Delta S_L(l)= \left\{ \begin{array}{lcc}
             \frac{c}{6}ln [2], & t \geq t_c, \\
             \ 0, & t < t_c.  
             \end{array}   
\right.
\label{stimator1}
\end{equation}
\noindent We expect that the behavior of $\Delta S$ will be  a step function as a function of  $t/U$. Even though other estimators have been proposed 
in the literature for determining the critical point using the block entropy~\cite{Xavier-PRB11}, we follow the L{\"a}uchli and Kollath proposal, 
because it works well for the Bose-Hubbard model.\par
\begin{figure}[t]
  \centering
\setlength{\fboxrule}{0.5 pt}
 \includegraphics [width=0.5\textwidth]{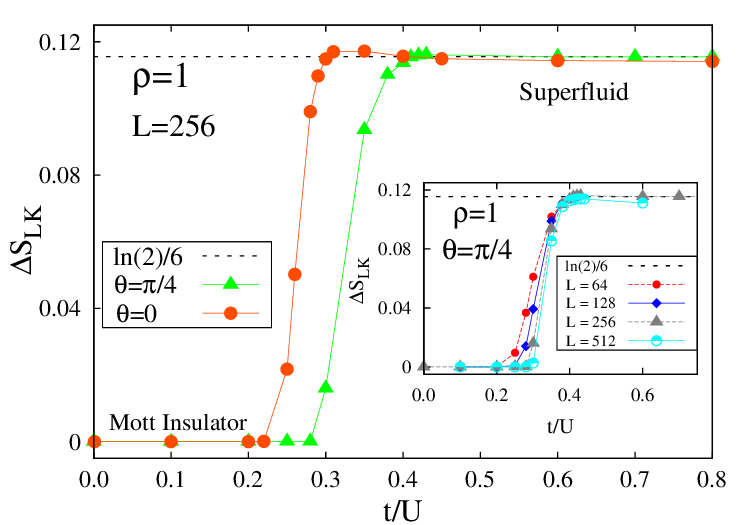} 
 \caption{ (Color online) The estimator $\Delta S_{LK}$ as a function of the hopping parameter t/U for $\theta=0$ and $\theta=\pi/4$. Here, we fixed 
 L=256 and $\rho=1$. In the inset, the estimator $\Delta S_{LK}$ vs $t/U$ for $\theta=\pi/4$, and different system lengths $L=64$, $128$, $256$, 
and $512$. The lines are visual guides.}
\label{fig5}
\end{figure}

 
In Fig. \ref{fig5}, we show the dependency between the estimator $\Delta S_{LK}$ and $t/U$ for $L=256$ and $\rho=1$ for two angles, $\theta=0$ and 
$\theta=\pi/4$.  The estimator is zero at $t=0$ for any value of the statistical angle and remains constant in a finite region. The width of this 
region depends on the statistical angle, i. e. the width of the Mott insulator area will increase with the angle $\theta$. The estimator grows 
quickly after a certain value of the hopping, which varies with the statistical angle and reaches the value $ln (2)/6$. This value corresponds to 
the estimator evaluated in a critical region according to the expression Eq. (\ref{stimator1}). Note that the estimator remains constant at this 
value for larger values of the hopping. It is clear from the figure that the anyon-Hubbard model exhibits a Mott insulator and a superfluid phase 
and that the behavior of the estimator corresponds to a step function. The value of the critical point is taken as the first value to reach 
$ln (2)/6$. In the inset of Fig. \ref{fig5}, we consider different system sizes, from 64 up to 512 sites, and observe that when the system size 
increases, the hopping for which $\Delta S_{LK}\neq 0$ moves to the right, and the curve tends to a step function as a function of $t/U$, and so 
we see that the expected behavior will occur.\par 
We found that the quantum critical point for $\theta=0$ and $\theta=\pi/4$ are $t_c/U=0.303$ and  $t_c/U=0.414$, respectively. Note that the result 
for the bosonic case $\theta=0$  is in accordance with previous results \cite{Ejima-EL11}. It is important to observe that the most accurate 
determination of the critical point gives us a bigger value in comparison with the calculation shown in Fig. \ref{fig3}, where we observe that the 
gap closes at $t/U=0.40$. The estimator results reinforce the idea that the inclusion of state-dependent hopping helps to localize the particles in 
this system; in other words, the inclusion of the correlated hopping means that the kinetic energy to delocate the system increases, since there is 
a displacement towards greater $t/U$ of the critical point.\par 

\begin{figure}[t]
  \centering
\setlength{\fboxrule}{0.5 pt}
 \includegraphics [width=0.5\textwidth]{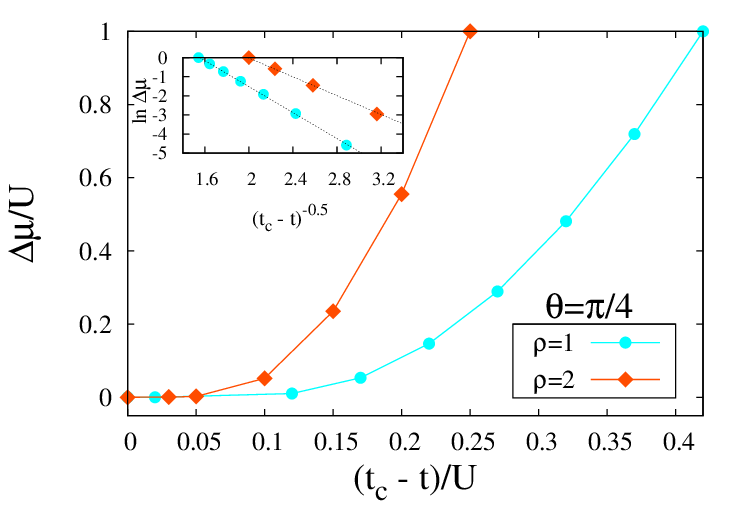} 
 \caption{ (Color online) Energy gap as a function of $t_c -t $ for $\rho=1$ and $\theta=\pi/4$. In the inset, $ln \hspace{0.1 cm} \Delta\mu$ 
 vs 1/$\sqrt{t_c -t}$. Here, the points are DMRG results, and the fits  to the Kosterlitz-Thouless transition are shown by lines.}
\label{fig6}
\end{figure}

We have characterized the quantum phases of the anyon-Hubbard model with $\theta=\pi/4$, and we have shown that the use of the L{\"a}uchli and Kollath 
estimator allows us to better find the $t_c/U$ position for the case $\rho=1$. On the other hand, we showed that the central charge $c$ in 
the critical phase is very close to $1$. However, the kind of transition that is taking place has not yet been discussed. It is important to remember 
that for a fixed integer number of particles, the Bose-Hubbard model belongs to the universality class of the XY model; hence the gap closes following 
a Kosterlitz-Thouless formula~\cite{Kosterlitz-JPC73}. We present the energy gap $\Delta\mu/U$ as a function of $(t_c-t)/U$  for anyon-Hubbard model 
with $\theta=\pi/4$ and for densities $\rho=1$ and $\rho=2$ in Fig. \ref{fig6}. We found the critical point for $\rho=2$ following the same procedure 
used to determine the critical point for $\rho=1$. Regardless of the density, we obtained that the gap exhibits a linear dependence for larger values 
of $(t_c-t)/U$; however, as the parameter diminishes, the gap decreases smoothly, and finally we observe that the gap vanishes very slowly, 
corroborating the results shown in Fig. \ref{fig3}. The above results obtained for the anyon-Hubbard model suggest that we try to fit the curves in 
Fig. \ref{fig6} to the Kosterlitz-Thouless formula:
\begin{equation}
\frac{\Delta\mu}{U}=\frac{\mu^p-\mu^h}{U}\sim exp \left(\frac{const.}{\sqrt{(t_c-t)/U}}\right). 
\label{ktt}
\end{equation}
In the inset of Fig. \ref{fig6}, we present $ln \hspace{0.1 cm}(\Delta\mu/U)$ as a function of 1$/\sqrt{(t_c-t)/U}$, which shows a linear dependence 
for both densities. Therefore, we affirmed that the transitions are of the Kosterlitz-Thouless kind for these densities under the parameters 
considered. The above result and the fact that the central charge is close to $1$ allow us to infer that the anyon-Hubbard model with $\theta=\pi/4$ 
is in the same universality class as the  Bose-Hubbard model.\par 
\begin{figure}[t]
\centering
\setlength{\fboxrule}{0.5 pt}
 \includegraphics [width=0.5\textwidth]{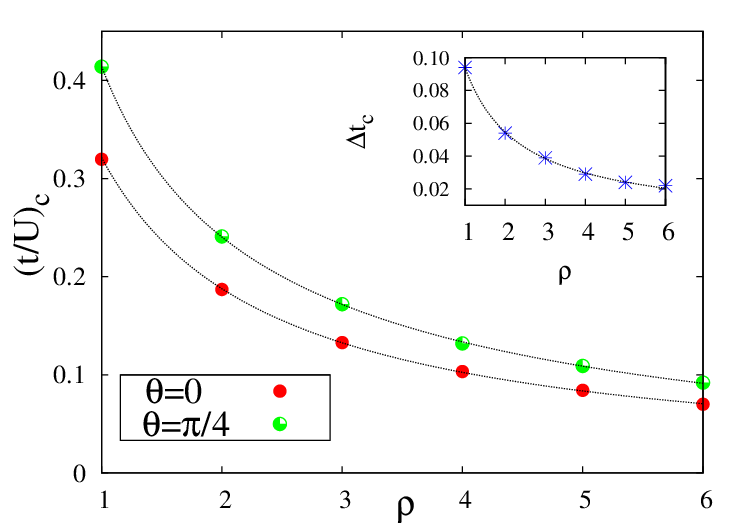} 
 \caption{ (Color online) Quantum critical point position $(t/U)_c$ as a function of the density for the anyon-Hubbard model with $\theta=\pi/4$ 
 and $\theta=0$. The dashed lines represent the best fit of the numerical data with the function of Eq. (\ref{arel}).  The numerical constants 
 obtained for $\theta=\pi/4$ are $(\alpha,\beta,\gamma)=(-0.037,0.45,-0.7)$  Inset: $\Delta t_c$ between $\theta=0$ and $\theta=\pi/4$ as a function 
 of the density $\rho$. (the data of $\theta=0$ were taken from \cite{Danshita-PRA11}) }
\label{fig7}
\end{figure}

In Fig. \ref{fig3}, we show that the position of the critical points moves to lower values as the density increases. It is noteworthy that the 
functional dependency of the critical points of the superfluid-to-Mott-insulator transition with the density for the Bose-Hubbard was a problem 
addressed by Danshita {\it et al.} in 1D, 2D, and 3D~\cite{Danshita-PRA11}. They found that the critical values versus the density are well 
approximated by the function $\left(\frac{U}{D\rho t}\right)_c=a+b\rho^{-c}$, where $D$ denotes the dimensionality of the system  and the 
constants $a,b$, and $c$ are numerically determined. We wanted to find out whether the expression obtained by Danshita {\it et al.} is valid for the 
anyon-Hubbard model with $\theta= \pi/4$, so we increased the local Hilbert space, considering $\rho+5$ states when the ground state 
with $\rho$ particles per site is taken into account. This correction allows us to determine the critical points with more precision; however, the 
computational cost increases. Using the  estimator (\ref{stimator1}) to find the critical points for higher densities, we obtain the results 
shown in Fig. \ref{fig7} for a fixed value of the statistical angle ($\theta=\pi/4$) and its comparison with the bosonic case $\theta=0$ 
(data taken from \cite{Danshita-PRA11}). We can see how as the density increases, the position of the critical point moves toward progressively 
smaller $t$ values, which implies that the insulator region decreases as we increase the filling factor of the system. Within the interval of 
densities studied, the curve for anyons is above the curve for bosons. This implies that the Mott lobe is always greater for the anyon case. 
Nevertheless, as we increase the density, the difference between the critical points decreases.\par 
The best fit of the numerical data of Fig. \ref{fig7} was obtained using the relation 
\begin{equation}
\left(\frac{t}{U}\right)_c=\alpha+\beta\rho^{-\gamma}, 
\label{arel}
\end{equation}
\noindent with $\alpha=-0.037$, $\beta=0.45$, and $\gamma=-0.7$ for the anyon case ($\theta=\pi/4$). Note that the above expression is different from 
the general formula found by Danshita {\it et al.}. We see in the inset of Fig. \ref{fig7} the difference between the critical point positions of 
the Bose- and anyon-Hubbard model. This quantity decreases  as the density grows and does not cancel out, which reflects the influence of the 
density-dependent hopping.\par  

\begin{figure}[t]
\centering
\setlength{\fboxrule}{0.5 pt}
 \includegraphics [width=0.5\textwidth]{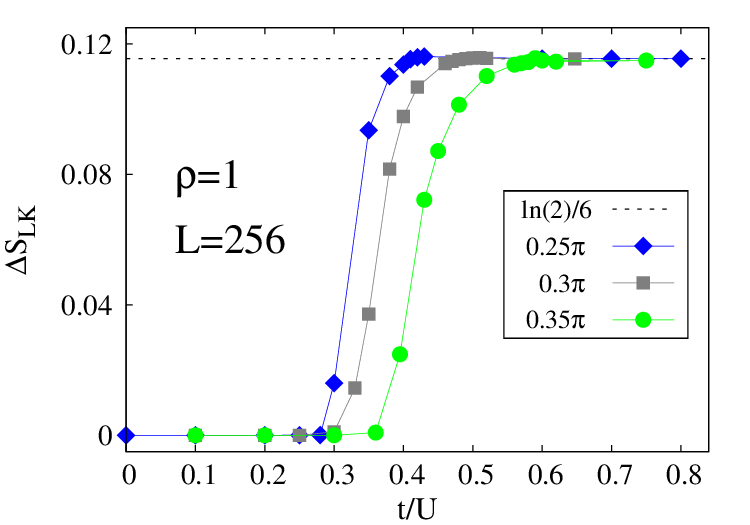} 
 \caption{  (Color online) The estimator $\Delta S_{LK}$ as a function of the hopping parameter $t/U$ for various statistical angles 
 $\theta=0.25\pi, 0.3\pi$ and $0.35\pi$. Here, we fixed L=256 and $\rho=1$. }
\label{fig8}
\end{figure}
Due to the above mentioned motivation, we concentrated our study of the anyon-Hubbard model on the statistical angle $\theta=\pi/4$, but now we want 
to explore the evolution of the critical point position as a function of the angle, and so we consider larger values of the statistical angle, as 
 is shown in Fig. \ref{fig8}. There, we present the results of the L{\"a}uchli and Kollath's estimator $\Delta S_{LK}$ as a function of $t/U$ for 
a system with one boson per site and three different values of the statistical angle were considered. Regardless of the $\theta$ value, the behavior 
of the estimator is similar. We observe that it is zero in a finite region of $t/U$ values, indicating that the system is in a Mott insulator state 
in this region. Note that the size of this region increases with the angle in a nonlinear proportion. After a certain value, which depends on the 
statistical angle, the estimator increases quickly and reaches the value of $(\text{ln} 2)/6$. For larger values of $t/U$, the estimator remains 
constant at the latter value, which indicates that the system is in a superfluid state, in accordance with the expression Eq. (\ref{stimator1}). 
Although the results shown in this figure correspond to a lattice size of $L=256$, we expected that for bigger lattices this behavior would be 
maintained, and a step function would be obtained in a manner similar to the inset of Fig. \ref{fig5}. On the other hand, we observe that the number 
of the states per block to reach the limit value  $(\text{ln} 2)/6$ increases dramatically with the statistical angle. From Fig. \ref{fig8}, we 
confirm that particles tend to localize when the statistical angle grows, a fact that is marked by an increase in the Mott insulator lobes area; 
therefore, the position of the critical point moves to larger values.\par
\begin{figure}[t]
  \centering
\setlength{\fboxrule}{0.5 pt}
 \includegraphics [width=0.5\textwidth]{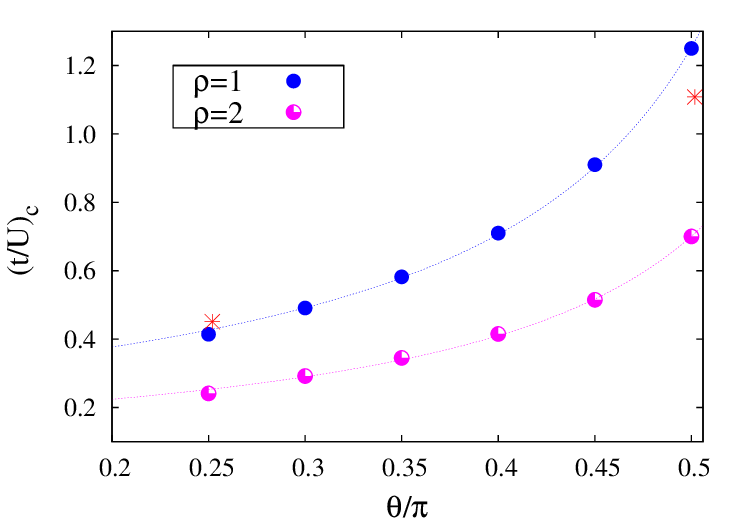} 
 \caption{ (Color online) Critical point evolution with statistical angle $\theta$ for the anyon-Hubbard model. The dashed lines represent the best 
fit of the numerical data. The stars represent the critical points found by Keilmann {\it et al.}. We can not explore larger values of $\theta$ due 
to the dramatic increase in the number of states that must be maintained in order to achieve the limit $(\text{ln} 2)/6$.} 
\label{fig9}
\end{figure}
Using the vanishing gap criteria, Keilmann {\it et al.} estimate the evolution of the critical points as a function of the statistical angle for a 
global density $\rho=1$, showing that critical strength $U_c/t$ decreases with $\theta$ and vanishes at $\theta=\pi$~\cite{Keilmann-NC11}. Today, we 
know that the above first approximation criterion is not very accurate, and taking into account that the L{\"a}uchli and Kollath estimator allows  
identify the border between the Mott insulator and the superfluid phases, we study the evolution of the critical points as the statistical angle 
increases in a chain with one or two particles per site (Fig. \ref{fig9}). We observe that the critical point increases gradually and smoothly 
with $\theta$, regardless of the global density $\rho$, which reflects the increase in the localization of the particles. The effect of the repulsion 
between the particles is evident in Fig. \ref{fig9}, since for $\rho=2$ more particles interact and the required kinetic energy to pass to the 
superfluid state is less than for the $\rho=1$ case for any value of $\theta$. Note that the position of the critical points for $\rho=2$ moves to 
greater values more slowly than in the $\rho=1$ case as the statistical angle increases. When we compare the position of the critical points for 
the first Mott lobe found by Keilmann {\it et al.} with our results, we observe that for small (large) angles  our critical point position moves to 
lower (greater) values compared to the Keilmann {\it et al.} results.\par


\subsection{\label{iv} Conclusions}

The ground state of the one-dimensional anyon-Hubbard model was studied, considering a mapping to a modified Bose-Hubbard model, which was explored using the 
density matrix renormalization group method. We found that the anyon-Hubbard model exhibits two quantum phases: Mott insulator and superfluid. We 
presented a phase diagram for anyons with $\theta=\pi/4$ for the first three densities ($\rho=1,2$, and $3$) and concluded that the density 
increase favors the appearance of the superfluid region, while the position of the critical point decreases. This result contradicts recently 
reported calculations of mean field theory and DMRG~\cite{Zhang-arxiv15}. A reentrance phase transition was observed for the three densities. We calculated the evolution 
of the critical points with the density for $\theta=\pi/4$ and found an analytical expression $t_c/U=-0.037+0.45\rho^{-0.7}$ using the von Neumann block 
entropy and the estimator proposed by L{\"a}uchli and Kollath. For $\rho=1$ and $\rho=2$, we showed that the gap closing can be fit to a 
Kosterlitz-Thouless expression, and taking into account that the central charge is $c=0.97$, we argued that the anyon-Hubbard model with 
$\theta=\pi/4$ belongs to the same universality class as the Bose-Hubbard model. We observed that the increase of the statistical angle leads to the 
localization of the particles, a fact that can be relevant when many-body interactions between particles are considered, because, for instance 
this can lead to obtaining Mott insulator phases for any density.\par  

\section*{Acknowledgments}
The authors are thankful for the support of DIB- Universidad Nacional de Colombia and COLCIENCIAS (grant No. FP44842-057-2015). Silva-Valencia and 
Franco are grateful for the hospitality of the ICTP, where part of this work was done.

\bibliography{Bibliografia}

\end{document}